\documentclass[twocolumn, prd, showpacs, showkeys, amsmath, amssymb, longbibliography, nofootinbib, floatfix, preprintnumbers]{revtex4-1}
\pdfoutput=1
\usepackage{graphicx}
\usepackage{enumerate}
\usepackage{bm}
\usepackage{slashed}
\usepackage{comment}
\usepackage{bbold}
\usepackage{braket}
\usepackage{makecell}
\usepackage{amssymb}
\usepackage{epsfig}
\usepackage{epstopdf}

\newcommand\Dslash {\slashed{D}}

\newcommand{\be}{\begin{equation}}
\newcommand{\ee}{\end{equation}}
\newcommand\beq{\begin{eqnarray}}
\newcommand\eeq{\end{eqnarray}} 
\newcommand\eqn[1]{\label{eq:#1}} 
\newcommand\eq[1]{eq.~(\ref{eq:#1})}

\newcommand{\vev}[1]{\langle #1 \rangle}

\newcommand{\bfdelt}{{\boldsymbol \delta}}

\newcommand{\GeV}{{\rm ~GeV }}
\newcommand{\TeV}{{\rm ~TeV }}

\newcommand{\CB}{{\cal B}}

\newcommand{\CO}{{\cal O}}

\newcommand{\CG}{{\cal G}}

\newcommand{\CT}{{\cal T}}

\newcommand{\CY}{{\cal Y}}

\newcommand{\CL}{{\cal L}}

\newcommand{\Tr}{{\rm Tr\,}}
\newcommand\half{{\textstyle{\frac{1}{2}}}}

\newcommand{\mybar}[1]%
        {\kern 0.6pt\overline{\kern -0.6pt#1\kern -0.6pt}\kern 0.6pt}
\makeatletter
\renewcommand*\env@matrix[1][\arraystretch]{%
  \edef\arraystretch{#1}%
  \hskip -\arraycolsep
  \let\@ifnextchar\new@ifnextchar
  \array{*\c@MaxMatrixCols c}}
\makeatother
\begin{document}

\preprint{INT-PUB-15-052}

\title{Little Flavor and $U(2)$ Family Symmetry}

\author{Dorota M. Grabowska}
 \email{grabow@uw.edu}

\author{David B. Kaplan}
\email{dbkaplan@uw.edu}
\affiliation{Institute for Nuclear Theory, Box 351550, Seattle, WA 98195-1550, USA}
 
 \date{\today}
 
 \begin{abstract}
 We examine an effective field theory inspired by Little Flavor that demonstrates a new paradigm for generating quark and lepton masses in which the scale of new flavor physics can be at the few TeV level, and new $Z'$ and $W'$ bosons are predicted. The model possesses an approximate $U(2)^2$ vector symmetry, not the full approximate $U(2)^5$ chiral symmetry of the Standard Model or Minimal Flavor Violation models, yet flavor changing neutral currents are sufficiently  suppressed. Additionally, lepton flavor violating processes, such as $\mu\to 3e$, lie naturally just below experimental bounds and the down quark mass can be radiatively generated.
  \end{abstract}

\pacs{
12.15.-y, 11.30.Hv, 12.60.-i
}
\maketitle
\section{Introduction}

The origin of flavor in the Standard Model (SM) remains one of the great mysteries of particle physics.  While patterns seem to exist in the quark masses and mixings, they have not led to any particularly convincing models for flavor. Furthermore, the absence of observed electric dipole moments,  highly suppressed rare decays, such as $\mu\to 3e$, and the small neutrino masses all provide circumstantial evidence that the next new physics may lie at extremely short distance scales, out of the reach of collider experiments, and quite possibly precision experiments.  However, the large suppression of flavor changing neutral currents (FCNC)  sends a more ambiguous message. It can certainly be explained by a dearth of new physics until high energy scales, but it could also be compatible with experimentally accessible physics provided there is some approximate symmetry that ensures that neutral currents are approximately flavor-diagonal in the mass eigenstate basis.   As this is exactly what suppresses FCNC in the SM, such a possibility may not be far-fetched. 

The operative symmetry in the SM and in some models of physics beyond the SM, such Minimal Flavor Violation (MFV) models, is an approximate chiral $U(2)^5$ symmetry acting on the lightest two families of quarks and leptons.  An interesting alternative, which occurs in Extended Technicolor and some supersymmetric models, is a much smaller vector $U(2)$ symmetry acting on the lightest two families of quarks and another $U(2)$ acting on the leptons.  Models with a vector $U(2)$ are more interesting phenomenologically as they typically allow for observable rare lepton decays, suppressed by the square of the muon mass, not by the much smaller neutrino masses. Such a vector symmetry does not by itself completely account for the smallness of FCNCs and an additional suppression mechanism must be present.  In this Letter we summarize how these symmetries are implemented, and, using effective field theory (EFT) techniques, show how a new paradigm for beyond the SM physics based on  Little Flavor~\cite{Sun:2013cza} provides a viable realization of approximate $U(2)^2$ flavor symmetry.

\section{Symmetries and FCNC}

The $U(2)^5$ approximate flavor symmetry of the SM corresponds to independent rotations between first- and second-family fermions in the five different SM representations $q,\ell,u^c,d^c,e^c$.  Restricting our analysis to a two-family version of the SM, this symmetry is broken by the Yukawa matrices which act as spurions transforming as 
\beq
Y_{u,d}\to U_q Y_{u,d} U_{u,d}^\dagger\ ,\qquad Y_{e}\to U_\ell Y_{e} U_{e}^\dagger\ .
\eeq
with $U_{i}$ being $U(2)$ matrices. The $\Delta S=2$ operator giving rise to $M_{12}$ in the kaon system comes from a one-loop diagram; 
working in the mass basis of the down quarks, $Y_d$ is diagonal and the operator must, by symmetry, be proportional to $ \left(Y_uY_u^\dagger\right)_{12}^2/16\pi^2 G_F(\mybar d_L \gamma^\mu s_L)^2$.  However, quarks become massless in the zero $Y_{u,d}$ limit and so the operator need not be analytic in $Y$. In fact the box diagram has IR singularities in that limit, and thus the prefactor is actually of the form  $(Y_uY_u^\dagger)_{12}^2/(16\pi^2\Tr Y_uY_u^\dagger)$. The dominant breaking of $U(2)^3$ symmetry in the quark sector is due to the charm mass, and so the SM result is $M_{12}\propto G_F^2 \sin^2\theta_c m_c^2$, where $\theta_c$ is the Cabbibo angle. In the realistic three family case the operative symmetry is still an approximate $U(2)^5$ and not $U(3)^5$, as the top is heavy but with small coupling to the first two families. Therefore, two additional doublet spurions account for the coupling of the third family to the first two, namely $\CT=\lambda_t\left\{V_{td}^*,V_{ts}^*\right\}$ and $\CB=\left\{\lambda_u V_{ub},\lambda_c V_{cb}\right\}$, where $\lambda_i$ are quark Yukawa couplings. This contribution to $M_{12}$ is proportional to $\left(\CT^\dagger\CT + \CB^\dagger \CB\right)_{12}$, which determines $\epsilon_K$ but is not the dominant contribution to $\Delta M_K$.

In Minimal Flavor Violation, one assumes that there are no new sources of flavor violation besides  the SM Yukawa couplings and that the effective theory also obeys an $U(2)^5$ symmetry \cite{Chivukula:1987py, D'Ambrosio:2002ex}. The contributions to $\Delta S = 2$ processes from new physics at scale $\Lambda$ is proportional to $(Y_u Y_u^\dagger)^2_{12}/\Lambda^2$.  With two families, $(Y_u Y_u^\dagger)^2_{12}/\Lambda^2 = m_c^4\theta_c^2/\Lambda$, which sets the bound $\Lambda\gtrsim\CO(10 \GeV)$.  With three families, the largest contribution is  $\lambda_t^4 V_{td}^{*2}V_{ts}^2/\Lambda^2$, which gives the much higher bound $\Lambda \ge \CO(5 \TeV)$. The MFV paradigm also gives bounds from rare lepton decays, once neutrino masses are introduced. For Dirac neutrinos, rare lepton decays are suppressed by at least two powers of the neutrino Yukawa coupling, in complete analogy with the quark sector. For masses generated via a seesaw mechanism, rare lepton decays are only observable if the scale of lepton number violation is vastly different than the scale of lepton flavor violation \cite{Cirigliano:2005ck}.

In Extended Technicolor (ETC), the approximate family symmetry is reduced to a vector-like $U(2)$ in the quark sector. This symmetry is gauged in order to generate SM fermion masses via radiative corrections. The mass of the fermions is given by $m_q = g^2_{ETC}\vev{\mybar{F}F}/2 M_{ETC}$ where $\vev{\mybar{F}F}$ is the condensate of the technifermions and $M_{ETC}$ is the mass of the gauge bosons that link SM fermions to technifermions. The gauged $U(2)$ also gives rise to tree level FCNC four-quark operators proportional to $\cos^2\theta\, g^2_{ETC}/\mybar M^2_{ETC}$, where $\theta$ is some mixing angle and $\mybar M_{ETC}$ is the mass of the gauge bosons that only mediate between SM fermions. Whereas the exact relative size of $M_{ETC}$ and $\mybar M_{ETC}$ depends on the details of ETC symmetry breaking, they will be on the same order\footnote{In a weakly coupled theory, group theoretic considerations allow for $M_{ETC}\gg \mybar M_{ETC}$ but not the reverse, since the generators associated with gauge bosons of mass $\mybar M_{ETC}$ can form a closed algebra, while those associated with $M_{ETC}$ cannot. Strong interactions have been invoked in Walking Technicolor \cite{Appelquist:1986an} to give rise to $M_{ETC}\ll \mybar M_{ETC}$, reconciling ETC with FCNC constraints.},
 which implies that the coefficient of the $\Delta S = 2$ operator can be no smaller than $\cos^2\theta\, m_c/\vev{\mybar{F}F}$. For small $\theta$, this is three orders of magnitude too large, and the theory requires an extra suppression mechanism, such as ``walking" \cite{Appelquist:1986an}.

In supersymmetric models, the vector $U(2)$ violating spurions that give rise the the SM quark masses also generate the squark mass differences, though not the flavor universal squark mass, $m_{\tilde{q}}$ \cite{Barbieri:1995uv, Pomarol:1995xc, Barbieri:1997tu, Barbieri:1998em}. Working in the mass basis of the down quarks, the $U(2)$ symmetry constrains the SUSY contribution to $M_{12}$ to be $(V\delta^2_{\tilde{q}}V^\dagger)(U\delta^2_{\tilde{q}}U^\dagger)/m^6_{\tilde{q}}$, where $\delta^2_{\tilde{q}}$ is the mass squared difference for first two families of squarks and $V, U$ are superfield rotation matrices that diagonalize the mass squared and trilinear terms. $V, U$ and $\delta^2_{\tilde{q}}$ are related to the spurions that generate the SM quark masses, and one finds the constraint of $m_{\tilde{q}}\gtrsim 2.5\TeV$ \cite{Hagelin:1992tc, Gabbiani:1996hi} is adequate to supply the needed  suppression.

Here we present an alternative method of introducing an approximate $U(2)$ horizontal symmetry that allows for FCNCs to be compatible with new TeV physics. It is based on a low energy effective description of Little Flavor~\cite{Sun:2013cza} and its novel feature is $U(2)$ flavor violating spurions with additional transformation properties under a nonlinearly realized symmetry of the Higgs sector. This setup allows for sufficiently suppressed FCNCs
 while also protecting the Higgs mass from large quadratic divergence, via the Composite Higgs \cite{Kaplan:1983fs,Kaplan:1983sm,Georgi:1984ef,Dugan:1984hq} and Little Higgs mechanisms \cite{ArkaniHamed:2002qy}. Note that the Little Higgs cancellations are incomplete in the effective theory considered here, where heavy fermions have been integrated out. The model predicts  rare leptonic decays that are suppressed by charged lepton masses, rather than the much smaller neutrino masses, and therefore potentially observable.

\section{Two families of quarks}
Consider a model with the following content:  two families of SM quarks $q_L^i$, $u_R^i$, $d_R^i$, $i=1,2$, and a $4\times 4$ unitary nonlinear sigma field $\Sigma$ which contains two composite Higgs doublets $H_{u,d}$ as well as other pseudo Goldstone bosons.  The theory has an approximate global symmetry
\beq
G_\text{global} &=& \left[SU(2)_L \times SU(2)_R  \times U(1)_{B-L} \times U(2)_f\right]\cr &&\times \left[SU(4)_{L} \times SU(4)_{R}\right]
\eeq
where the subgroup in the first bracket acts on the quarks, $U(2)_f$ is a horizontal family symmetry, and the group in the second bracket acts on the $\Sigma$ field. We will ignore color and the anomalous $U(1)_A$ symmetry, which play no role in this model.  We gauge a  $\CG_a\times \CG_b $ subgroup of this global symmetry, where  $\CG_{a,b} $ are independent $ SU(2)\times U(1) $ groups embedded in the global symmetry as
\beq
SU(2)_a &\subset& SU(2)_L  \times SU(4)_{L}\ ,\cr
U(1)_a &\subset& SU(2)_R \times U(1)_{B-L}\times  SU(4)_{L}\ ,\cr
SU(2)_b \times U(1)_b&\subset&  SU(4)_{R}
\eeq
The quarks have the usual SM quantum numbers under $SU(2)_a\times U(1)_a$
\beq
q^i_L = 2_{\frac{1}{6} } \ , \quad u^i_R = 1_{\frac{2}{3}}\ ,\quad d^i_R = 1_{-\frac{1}{3}}\ ,
\eeq
and are neutral under $SU(2)_b\times U(1)_b$.  The gauge generators act on the $\Sigma$ field as $\delta\Sigma =  iX_a\Sigma - i \Sigma X_b$. Denoting the four $X$ generators as  $\{gT^i,\,g'Y\}$ for both gauge groups, their representations on the $\Sigma$ field are given by
\beq
T^i =\half  \begin{pmatrix} \sigma_i &\\ & 0\end{pmatrix},\quad Y= \half\begin{pmatrix}0 &\\ & \sigma_3\end{pmatrix}\ .
\eeq
The $X$ generators act as spurions breaking the global symmetry, with $X_a$ transforming as $(\text{adjoint}\times 1) \oplus (1\times \text{adjoint})$ under the $\left[SU(2)_L\times SU(2)_R\right] \times SU(4)_L$ symmetry, and $X_b$ transforming as the adjoint under $SU(4)_R$. Before the introduction of additional spurions, the EFT is quite simple: the usual gauged chiral Lagrangian for $\Sigma$, quark kinetic terms, and fermion-meson interactions, which are $\CO(\alpha)$, that are required as counterterms.  The chiral Lagrangian is characterized by decay constant $f$ and cutoff $\Lambda\sim 4\pi f$; we will take $f\gtrsim 1.5 \TeV$.

If $\vev{\Sigma} = \mathbb{1}$, the $\left[SU(2)\times U(1)\right]^2$ gauge symmetry is spontaneously broken to the diagonal group $SU(2)\times U(1)$, which is identified with the SM gauge group. However we will assume the Composite Higgs paradigm that the vacuum misaligns and prefers the less symmetric ground state
\beq
\vev{\Sigma} = \begin{pmatrix}c_u&&s_u&\\&c_d&&s_d\\-s_u&&c_u&\\&-s_d&&c_d\end{pmatrix} \qquad \begin{array}{cc}c_i = \cos \frac{v_i}{f}\\  \\s_i = \sin \frac{v_i}{f}\end{array}
\eqn{vac}\eeq
which breaks the SM gauge group in the conventional way for a two Higgs doublet model; $v_i$ are the vevs of two Higgs doublets. The SM gauge couplings $g,g'$ are related to the $\CG_a \times \CG_b$ couplings by
\beq
g_a = \frac{g}{\cos\gamma_2},\,\,g'_a = \frac{g'}{\cos\gamma_1},\, \, g_b = \frac{g}{\sin\gamma_2},\, \,g'_b = \frac{g'}{\sin\gamma_1}
\eeq
and the exotic gauge bosons have TeV scales masses
\beq
M_{Z'}= M_{W'} = g f \csc 2\gamma_2 \quad M_{Z''} = g' f \csc 2\gamma_1
\eeq
with freedom to set the angles $\gamma_{1,2}$. We will constrain the gauge bosons masses once we add leptons.


\begin{table}[t!]
  \begin{center}
  \begin{tabular}{|l|ccccc|}
    \hline
 & $SU(2)_L$ & $SU(2)_R $& $U(2)_f$& $SU(4)_L $&$ SU(4)_R$  \\ \Xhline{2\arrayrulewidth}
$\CY_L$ & 2 &1 & 1& 4& 1\\ \hline
$\CY_R$ & 1 & 2& 1& 4&1 \\ \hline
$\Delta$ & 1 & 1 &  1$\,\oplus\,$3 & 1 & 15\\ \hline
$\epsilon$ & 1 & 1&  1$\,\oplus\,$3 & 10 & 1\\ \hline
 \end{tabular}
  \end{center}
  \caption{Transformation properties of spurions under the global approximate symmetry, where $\CY_{L,R}$ are given in \eq{Yspur} and $\epsilon$ in \eq{epsspur}.}
\label{tab:spurions}
\end{table}

Up to this point quark masses cannot be generated, even with $SU(2)\times U(1)$ breaking, as the quarks and Higgs transform under independent symmetries. We now assume additional particles that couple to the quark and $\Sigma$ fields, such as the heavy fermions of Ref.~\cite{Sun:2013cza}, have been integrated out at a scale $M<\Lambda$. This introduces additional spurions $\CY_{L,R}$  into the theory, which transform as shown in Table~\ref{tab:spurions}. These spurions take the form of rectangular $4\times 2$ matrices both transforming on the left by $SU(4)_L$ of the Higgs sector, and on right by either $SU(2)_L$ ($\CY_L$) or $SU(2)_R$ ($\CY_{R}$) of the quark sector:
\beq
\CY_L =\begin{pmatrix}  y_q & 0\\0& y_q\\ 0 & 0\\0&0 \end{pmatrix},\ 
\CY_R = \begin{pmatrix} 0 & 0\\0&0\\ y_u & 0\\0&y_d \end{pmatrix},\ 
\eqn{Yspur}\eeq
where $y < M/\Lambda <1$. $M$ also has to be larger than the heaviest bosons and so we take $M = \CO(5\TeV)$.   At tree level, matching operators proportional to $y^2 \Lambda^2/M^2$ will be induced in the EFT below the scale $M$. In operators proportional to $y^2$, derivatives appear in the ratio $D/M$ \footnote{Radiative corrections within the EFT below $M$ do not yield higher powers of $\Lambda/M$, since the cutoff in this theory is $M$ and the scale $\Lambda$ only appears in inverse powers as $1/\Lambda \sim 1/4\pi f$.}. To simplify the analysis, we assume $y\ll M/\Lambda$ and work to quadratic order in $y$; without this assumption, the bounds on the model are much less straightforward and thus we will leave it for a later paper.  Assume first that the heavy particles all have a universal mass $M$, invariant under $SU(4)_b\times U(2)_f$. By writing the RH quark fields  as a doublet, $q_R = \{u_R,d_R\}$, we can now add to the EFT  the operators
\beq
\CL_{y^2} =\frac{ \Lambda^2}{M^2} \left[ \mybar{q}_L \CY_L^\dagger  \Sigma \Dslash\Sigma^\dagger  \CY_L \,q_L+ L\leftrightarrow R \right]
\eqn{y2}\eeq
plus operators with two or more derivatives. For clarity, here and throughout, we neglect to write the $\CO(1)$ model-dependent Wilson coefficients.  This term contains new couplings of the quarks to SM and exotic gauge bosons. These new couplings between quarks and SM gauge bosons result in non-canonical kinetic terms and will require a finite wavefunction renormalization. 

The lowest dimension operator that couples $q_L$ to $q_R$ requires at least two derivatives and is of the form $\mybar q_L \CY_L^\dagger (\Sigma D^2\Sigma^\dagger)\CY_R q_R$. Hence, despite the fact that $\Sigma$ breaks the weak interactions, there are no quark mass terms at $\CO(y^2)$. The necessity of $SU(4)_b$ violation to generate fermion masses is the essence of the Little Flavor mechanism. The leading contributions appear at $\CO(y^2\alpha_b/4\pi)$ and give a $U(2)_f$ invariant common mass to the quarks:
\beq
\CL_{y^2\alpha_b}  = \frac{ \Lambda^2}{M} \frac{\alpha^i_b}{4\pi}  \mybar{q}_L \CY_L^\dagger \left(\Sigma X_b^i X_b^i\Sigma^\dagger\right) \CY_R \,q_R+ \text{h.c.}
\eqn{RadMass}
\eeq
where $i$ is summed over $0,\ldots,3$ with $i=0$ corresponding to the $U(1)_b$ and $i=1,2,3$ corresponding to $SU(2)_b$. 

Before analyzing the size of this mass term, let us first discuss the generation of nondegenerate masses. We drop the assumption that the heavy particles have an universal mass $M$, and replace $M\to M(1-\Delta)$ everywhere with 
\beq
\Delta = \begin{pmatrix}\bfdelt_u+ \bfdelt_d&&&\\& \bfdelt_u+ \bfdelt_d&&\\&& \bfdelt_d-3 \bfdelt_u&\\&&& \bfdelt_u- 3 \bfdelt_d\end{pmatrix}
\eeq
where $SU(4)_R$ space is shown, and $\bfdelt_{u,d}$ are two independent matrices in the flavor $U(2)_f$ space. We treat $\Delta$ as a small spurion transforming as in Table~\ref{tab:spurions}. Since it transforms nontrivially under $SU(4)_R$ it is now possible to construct mass terms at $\CO(y_Ly_R\Delta)$ of the form
\beq
\CL_{y^2\Delta} =   \frac{ \Lambda^2}{M} \mybar{q}_L \CY_L^\dagger \Sigma\Delta \Sigma^\dagger \CY_R \,q_R+ \text{h.c.}
\eqn{mass2}\eeq
The up-type quark mass matrix is then given by the sum of terms
\beq
M_u = -v_u\frac{\eta_q\eta_uM}{f}\left( \frac{3\alpha_b - \alpha'_b}{16\pi}+ 4\bfdelt_u\right)
 \eeq
where $\eta_{i} \equiv y_{i}\Lambda/M$ and the sign of $\Delta$ determines the relative sign between the two contributions. The down-type quark mass matrix is the same  with subscripts $u,d$ interchanged. To avoid fine tuning, the maximal size of the universal $\CO(y^2\alpha_b/4\pi)$ term  should be $\CO(m_u)$. Therefore, the charm mass arises mainly from the $\bfdelt_u$ term, as $m_c\gg m_u$, and so one of the eigenvalues of $4\eta_q\eta_u M \bfdelt_u/f$ must be $\lambda_c$. Since $\Delta$ has to be perturbative ($3\bfdelt_{u,d} < 1$), $\gamma_{1,2}$ are bounded via
\beq
\frac{1}{3}\gtrsim \frac{\lambda_c}{\lambda_u}\frac{3\alpha_b - \alpha'_b}{64 \pi}\ .
\eeq
This bound is easier to satisfy in the down sector than in the up sector, as the amount of fine tuning is determined by the mass splitting. By having the two contributions partially cancel each other in the up sector, we can account for the physical up quark mass at the cost of moderate fine tuning.

Introducing $\Delta$ into single derivative operators, 
\beq
\CL_{y^2\Delta D} =  \frac{ \Lambda^2}{M^2} \mybar{q}_L \CY_L^\dagger \Sigma \Delta\Dslash\Sigma^\dagger  \CY_L \,q_L+ L\leftrightarrow R + \text{h.c.}\,\,\,
\eeq
leads to flavor off-diagonal couplings for the exotic gauge bosons. Focusing on the neutral $\Delta S = 1$ currents, to leading order in $(v/f)^2$ these have the form
\beq
&&\CL_{\Delta S =1} =  \frac{\eta^2_d M_{Z''}}{f}\mybar d_R \slashed{Z}'' \left[R_d^\dagger \left(\bfdelt_u-3\bfdelt_d\right)R_d \right]_{12}s_R \nonumber \\
&&\quad+  \frac{\eta^2_q M_{Z'}}{f}\mybar d_L \slashed{Z}' \left[L_d^\dagger \left(\bfdelt_u+\bfdelt_d\right)L_d \right]_{12}s_L + \text{h.c.}
\eeq
We can then bound the $\Delta S = 2$ currents from above 
\beq
\CL_{\Delta S=2}\lesssim \frac{\eta^2_q \lambda^2_c}{4 \eta^2_uM^2}\left[\left(\mybar d_L \gamma_\mu s_L\right)^2+\frac{\eta^4_d}{\eta_q^4}\left(\mybar d_R \gamma_\mu s_R\right)^2\right]
\eeq
where the bound is saturated when assuming all relevant mixing angles are $\CO(1)$.  If we also assume $\eta_q\sim\eta_{u,d}$, then we require $M\gtrsim10 \TeV$.

\section{Adding Leptons}
Dirac neutrino masses are generated in the same way as quark masses. This allows us to constrain these models with respect to $Z'$ phenomenology and rare leptonic decays. The universal $\CO(\alpha_b)$ lepton mass can be made  small without any fine-tuning, unlike in the quark sector, as the mass ratio $\mu/e$ is significantly smaller than $c/u$. 

To estimate the bounds on the $M_{Z'}$ and $M_{Z''}$, we rescale the bounds on $Z'_{SSM}$\cite{Aad:2014cka}, where the Sequential Standard Model (SSM) assumes that the $Z'_{SSM}$ has the same gauge coupling at the SM $Z$ boson. For $\eta_{\ell, \nu, e}\ll1$ the $Z'$ and $Z''$ couplings are $g \tan \gamma_2 T_3$ and $g' \tan \gamma_1 Q$, respectively, where $T_3$ is the neutral gauge generator of the SM $SU(2)$ and $Q$ is the electric charge generator. Taking $f= 1.5\TeV$, the mass bounds are
\beq
M_{Z'}\gtrsim 2.0 \TeV \quad M_{Z''}\gtrsim 1.6 \TeV \quad \gamma_{1,2}\leq \frac{\pi}{4}\,\,\, \phantom{3}
\eeq
For rare lepton decays, the neutral currents that change lepton flavor are completely analogous to the $\Delta F = 1$ currents in the quark sector. The branching fraction, $\mathbf{B}$, for $\mu \rightarrow 3e$ can be bounded from above by:
\beq
\mathbf{B}_{\mu \rightarrow 3 e}\lesssim\frac{\lambda^2_\mu}{2 G_F^2 f^2 M^2}\left(\frac{45 \eta^2_e\sin^4\gamma_1 }{\eta_l^2}+ \frac{\eta^2_l\sin^4\gamma_2}{\eta_e^2}\right)\,\,\,\,
\eeq
where the bound is saturated for $\CO(1)$ mixing angles and we assume the neutrino contribution is negligible. For $\gamma_{1,2}$ that saturate the bounds on $M_{Z'}$ and $M_{Z''}$ and $M \sim 5\TeV$ the branching ratio is a tenth of the experimental bound, if $\eta_l\sim \eta_e$. It is worth noting that the large mass scale $M$ in the lepton sector should be on the same order as the mass scale in the quark sector.

It is interesting to consider Majorana neutrinos instead of Dirac and implement a seesaw mechanism. There is a Majorana mass term at  
$\CO(y^2 \Delta^2\epsilon)$ of the form
\beq
\CL_{\epsilon y^2\Delta^2}&=&  \frac{\Lambda^3}{ M^2}\,\left[\left(\Sigma \Delta \Sigma^\dagger\right)  \CY_L \ell  \right]^T \epsilon  \left[\left(\Sigma \Delta \Sigma^\dagger\right)  \CY_L \ell \right] + \text{h.c.}\ ,\cr
\epsilon &=&\frac{\Lambda}{M_N} \begin{pmatrix}0&0&0&0\\0&0&0&0\\ 0&0&\bfdelt_\nu & 0\\ 0&0&0&0\end{pmatrix}\ 
\eqn{epsspur}\eeq
where $\epsilon$ is a new $\Delta L=2$ spurion transforming as in Table~\ref{tab:spurions} and $M_N$ is the large Majorana mass.  There are additional contributions with $\Delta$ replaced by gauge generators. The analysis of a model with Majorana neutrinos is more complicated and we leave it for a future paper, along with a full treatment of three families and $CP$ violation. Majorana neutrinos were also the subject of Ref.~\cite{Sun:2014jha}.

%
\begin{acknowledgments}
We gratefully acknowledge
much useful guidance from 
Ann E. Nelson. This   work was supported by the National Science Foundation Graduate Research Fellowship under Grant No. DGE-1256082 and   by the DOE Grant No. DE-FG02-00ER41132.
\end{acknowledgments}

\sloppy
\bibliography{flavorbib}

\begin{thebibliography}{18}%
\makeatletter
\providecommand \@ifxundefined [1]{%
 \@ifx{#1\undefined}
}%
\providecommand \@ifnum [1]{%
 \ifnum #1\expandafter \@firstoftwo
 \else \expandafter \@secondoftwo
 \fi
}%
\providecommand \@ifx [1]{%
 \ifx #1\expandafter \@firstoftwo
 \else \expandafter \@secondoftwo
 \fi
}%
\providecommand \natexlab [1]{#1}%
\providecommand \enquote  [1]{``#1''}%
\providecommand \bibnamefont  [1]{#1}%
\providecommand \bibfnamefont [1]{#1}%
\providecommand \citenamefont [1]{#1}%
\providecommand \href@noop [0]{\@secondoftwo}%
\providecommand \href [0]{\begingroup \@sanitize@url \@href}%
\providecommand \@href[1]{\@@startlink{#1}\@@href}%
\providecommand \@@href[1]{\endgroup#1\@@endlink}%
\providecommand \@sanitize@url [0]{\catcode `\\12\catcode `\$12\catcode
  `\&12\catcode `\#12\catcode `\^12\catcode `\_12\catcode `\%12\relax}%
\providecommand \@@startlink[1]{}%
\providecommand \@@endlink[0]{}%
\providecommand \url  [0]{\begingroup\@sanitize@url \@url }%
\providecommand \@url [1]{\endgroup\@href {#1}{\urlprefix }}%
\providecommand \urlprefix  [0]{URL }%
\providecommand \Eprint [0]{\href }%
\providecommand \doibase [0]{http://dx.doi.org/}%
\providecommand \selectlanguage [0]{\@gobble}%
\providecommand \bibinfo  [0]{\@secondoftwo}%
\providecommand \bibfield  [0]{\@secondoftwo}%
\providecommand \translation [1]{[#1]}%
\providecommand \BibitemOpen [0]{}%
\providecommand \bibitemStop [0]{}%
\providecommand \bibitemNoStop [0]{.\EOS\space}%
\providecommand \EOS [0]{\spacefactor3000\relax}%
\providecommand \BibitemShut  [1]{\csname bibitem#1\endcsname}%
\let\auto@bib@innerbib\@empty
\bibitem [{\citenamefont {Sun}\ \emph {et~al.}(2013)\citenamefont {Sun},
  \citenamefont {Kaplan},\ and\ \citenamefont {Nelson}}]{Sun:2013cza}%
  \BibitemOpen
  \bibfield  {author} {\bibinfo {author} {\bibfnamefont {Sichun}\ \bibnamefont
  {Sun}}, \bibinfo {author} {\bibfnamefont {David~B.}\ \bibnamefont {Kaplan}},
  \ and\ \bibinfo {author} {\bibfnamefont {Ann~E.}\ \bibnamefont {Nelson}},\
  }\bibfield  {title} {\enquote {\bibinfo {title} {{Little flavor: A model of
  weak-scale flavor physics}},}\ }\href {\doibase 10.1103/PhysRevD.87.125036}
  {\bibfield  {journal} {\bibinfo  {journal} {Phys.Rev.}\ }\textbf {\bibinfo
  {volume} {D87}},\ \bibinfo {pages} {125036} (\bibinfo {year} {2013})},\
  \Eprint {http://arxiv.org/abs/1303.1811} {arXiv:1303.1811 [hep-ph]}
  \BibitemShut {NoStop}%
\bibitem [{\citenamefont {Chivukula}\ and\ \citenamefont
  {Georgi}(1987)}]{Chivukula:1987py}%
  \BibitemOpen
  \bibfield  {author} {\bibinfo {author} {\bibfnamefont {R.~Sekhar}\
  \bibnamefont {Chivukula}}\ and\ \bibinfo {author} {\bibfnamefont {Howard}\
  \bibnamefont {Georgi}},\ }\bibfield  {title} {\enquote {\bibinfo {title}
  {{Composite Technicolor Standard Model}},}\ }\href {\doibase
  10.1016/0370-2693(87)90713-1} {\bibfield  {journal} {\bibinfo  {journal}
  {Phys.Lett.}\ }\textbf {\bibinfo {volume} {B188}},\ \bibinfo {pages} {99}
  (\bibinfo {year} {1987})}\BibitemShut {NoStop}%
\bibitem [{\citenamefont {D'Ambrosio}\ \emph {et~al.}(2002)\citenamefont
  {D'Ambrosio}, \citenamefont {Giudice}, \citenamefont {Isidori},\ and\
  \citenamefont {Strumia}}]{D'Ambrosio:2002ex}%
  \BibitemOpen
  \bibfield  {author} {\bibinfo {author} {\bibfnamefont {G.}~\bibnamefont
  {D'Ambrosio}}, \bibinfo {author} {\bibfnamefont {G.~F.}\ \bibnamefont
  {Giudice}}, \bibinfo {author} {\bibfnamefont {G.}~\bibnamefont {Isidori}}, \
  and\ \bibinfo {author} {\bibfnamefont {A.}~\bibnamefont {Strumia}},\
  }\bibfield  {title} {\enquote {\bibinfo {title} {{Minimal flavor violation:
  An Effective field theory approach}},}\ }\href {\doibase
  10.1016/S0550-3213(02)00836-2} {\bibfield  {journal} {\bibinfo  {journal}
  {Nucl. Phys.}\ }\textbf {\bibinfo {volume} {B645}},\ \bibinfo {pages}
  {155--187} (\bibinfo {year} {2002})},\ \Eprint
  {http://arxiv.org/abs/hep-ph/0207036} {arXiv:hep-ph/0207036 [hep-ph]}
  \BibitemShut {NoStop}%
\bibitem [{\citenamefont {Cirigliano}\ \emph {et~al.}(2005)\citenamefont
  {Cirigliano}, \citenamefont {Grinstein}, \citenamefont {Isidori},\ and\
  \citenamefont {Wise}}]{Cirigliano:2005ck}%
  \BibitemOpen
  \bibfield  {author} {\bibinfo {author} {\bibfnamefont {Vincenzo}\
  \bibnamefont {Cirigliano}}, \bibinfo {author} {\bibfnamefont {Benjamin}\
  \bibnamefont {Grinstein}}, \bibinfo {author} {\bibfnamefont {Gino}\
  \bibnamefont {Isidori}}, \ and\ \bibinfo {author} {\bibfnamefont {Mark~B.}\
  \bibnamefont {Wise}},\ }\bibfield  {title} {\enquote {\bibinfo {title}
  {{Minimal flavor violation in the lepton sector}},}\ }\href {\doibase
  10.1016/j.nuclphysb.2005.08.037} {\bibfield  {journal} {\bibinfo  {journal}
  {Nucl. Phys.}\ }\textbf {\bibinfo {volume} {B728}},\ \bibinfo {pages}
  {121--134} (\bibinfo {year} {2005})},\ \Eprint
  {http://arxiv.org/abs/hep-ph/0507001} {arXiv:hep-ph/0507001 [hep-ph]}
  \BibitemShut {NoStop}%
\bibitem [{\citenamefont {Appelquist}\ \emph {et~al.}(1986)\citenamefont
  {Appelquist}, \citenamefont {Karabali},\ and\ \citenamefont
  {Wijewardhana}}]{Appelquist:1986an}%
  \BibitemOpen
  \bibfield  {author} {\bibinfo {author} {\bibfnamefont {Thomas~W.}\
  \bibnamefont {Appelquist}}, \bibinfo {author} {\bibfnamefont {Dimitra}\
  \bibnamefont {Karabali}}, \ and\ \bibinfo {author} {\bibfnamefont {L.~C.~R.}\
  \bibnamefont {Wijewardhana}},\ }\bibfield  {title} {\enquote {\bibinfo
  {title} {{Chiral Hierarchies and the Flavor Changing Neutral Current Problem
  in Technicolor}},}\ }\href {\doibase 10.1103/PhysRevLett.57.957} {\bibfield
  {journal} {\bibinfo  {journal} {Phys. Rev. Lett.}\ }\textbf {\bibinfo
  {volume} {57}},\ \bibinfo {pages} {957} (\bibinfo {year} {1986})}\BibitemShut
  {NoStop}%
\bibitem [{\citenamefont {Barbieri}\ \emph {et~al.}(1996)\citenamefont
  {Barbieri}, \citenamefont {Dvali},\ and\ \citenamefont
  {Hall}}]{Barbieri:1995uv}%
  \BibitemOpen
  \bibfield  {author} {\bibinfo {author} {\bibfnamefont {Riccardo}\
  \bibnamefont {Barbieri}}, \bibinfo {author} {\bibfnamefont {G.~R.}\
  \bibnamefont {Dvali}}, \ and\ \bibinfo {author} {\bibfnamefont {Lawrence~J.}\
  \bibnamefont {Hall}},\ }\bibfield  {title} {\enquote {\bibinfo {title}
  {{Predictions from a U(2) flavor symmetry in supersymmetric theories}},}\
  }\href {\doibase 10.1016/0370-2693(96)00318-8} {\bibfield  {journal}
  {\bibinfo  {journal} {Phys. Lett.}\ }\textbf {\bibinfo {volume} {B377}},\
  \bibinfo {pages} {76--82} (\bibinfo {year} {1996})},\ \Eprint
  {http://arxiv.org/abs/hep-ph/9512388} {arXiv:hep-ph/9512388 [hep-ph]}
  \BibitemShut {NoStop}%
\bibitem [{\citenamefont {Pomarol}\ and\ \citenamefont
  {Tommasini}(1996)}]{Pomarol:1995xc}%
  \BibitemOpen
  \bibfield  {author} {\bibinfo {author} {\bibfnamefont {Alex}\ \bibnamefont
  {Pomarol}}\ and\ \bibinfo {author} {\bibfnamefont {Daniele}\ \bibnamefont
  {Tommasini}},\ }\bibfield  {title} {\enquote {\bibinfo {title} {{Horizontal
  symmetries for the supersymmetric flavor problem}},}\ }\href {\doibase
  10.1016/0550-3213(96)00074-0} {\bibfield  {journal} {\bibinfo  {journal}
  {Nucl. Phys.}\ }\textbf {\bibinfo {volume} {B466}},\ \bibinfo {pages} {3--24}
  (\bibinfo {year} {1996})},\ \Eprint {http://arxiv.org/abs/hep-ph/9507462}
  {arXiv:hep-ph/9507462 [hep-ph]} \BibitemShut {NoStop}%
\bibitem [{\citenamefont {Barbieri}\ \emph {et~al.}(1997)\citenamefont
  {Barbieri}, \citenamefont {Hall},\ and\ \citenamefont
  {Romanino}}]{Barbieri:1997tu}%
  \BibitemOpen
  \bibfield  {author} {\bibinfo {author} {\bibfnamefont {Riccardo}\
  \bibnamefont {Barbieri}}, \bibinfo {author} {\bibfnamefont {Lawrence~J.}\
  \bibnamefont {Hall}}, \ and\ \bibinfo {author} {\bibfnamefont {Andrea}\
  \bibnamefont {Romanino}},\ }\bibfield  {title} {\enquote {\bibinfo {title}
  {{Consequences of a U(2) flavor symmetry}},}\ }\href {\doibase
  10.1016/S0370-2693(97)00372-9} {\bibfield  {journal} {\bibinfo  {journal}
  {Phys. Lett.}\ }\textbf {\bibinfo {volume} {B401}},\ \bibinfo {pages}
  {47--53} (\bibinfo {year} {1997})},\ \Eprint
  {http://arxiv.org/abs/hep-ph/9702315} {arXiv:hep-ph/9702315 [hep-ph]}
  \BibitemShut {NoStop}%
\bibitem [{\citenamefont {Barbieri}\ \emph {et~al.}(1999)\citenamefont
  {Barbieri}, \citenamefont {Giusti}, \citenamefont {Hall},\ and\ \citenamefont
  {Romanino}}]{Barbieri:1998em}%
  \BibitemOpen
  \bibfield  {author} {\bibinfo {author} {\bibfnamefont {Riccardo}\
  \bibnamefont {Barbieri}}, \bibinfo {author} {\bibfnamefont {Leonardo}\
  \bibnamefont {Giusti}}, \bibinfo {author} {\bibfnamefont {Lawrence~J.}\
  \bibnamefont {Hall}}, \ and\ \bibinfo {author} {\bibfnamefont {Andrea}\
  \bibnamefont {Romanino}},\ }\bibfield  {title} {\enquote {\bibinfo {title}
  {{Fermion masses and symmetry breaking of a U(2) flavor symmetry}},}\ }\href
  {\doibase 10.1016/S0550-3213(99)00195-9} {\bibfield  {journal} {\bibinfo
  {journal} {Nucl. Phys.}\ }\textbf {\bibinfo {volume} {B550}},\ \bibinfo
  {pages} {32--40} (\bibinfo {year} {1999})},\ \Eprint
  {http://arxiv.org/abs/hep-ph/9812239} {arXiv:hep-ph/9812239 [hep-ph]}
  \BibitemShut {NoStop}%
\bibitem [{\citenamefont {Hagelin}\ \emph {et~al.}(1994)\citenamefont
  {Hagelin}, \citenamefont {Kelley},\ and\ \citenamefont
  {Tanaka}}]{Hagelin:1992tc}%
  \BibitemOpen
  \bibfield  {author} {\bibinfo {author} {\bibfnamefont {John~S.}\ \bibnamefont
  {Hagelin}}, \bibinfo {author} {\bibfnamefont {S.}~\bibnamefont {Kelley}}, \
  and\ \bibinfo {author} {\bibfnamefont {Toshiaki}\ \bibnamefont {Tanaka}},\
  }\bibfield  {title} {\enquote {\bibinfo {title} {{Supersymmetric flavor
  changing neutral currents: Exact amplitudes and phenomenological
  analysis}},}\ }\href {\doibase 10.1016/0550-3213(94)90113-9} {\bibfield
  {journal} {\bibinfo  {journal} {Nucl. Phys.}\ }\textbf {\bibinfo {volume}
  {B415}},\ \bibinfo {pages} {293--331} (\bibinfo {year} {1994})}\BibitemShut
  {NoStop}%
\bibitem [{\citenamefont {Gabbiani}\ \emph {et~al.}(1996)\citenamefont
  {Gabbiani}, \citenamefont {Gabrielli}, \citenamefont {Masiero},\ and\
  \citenamefont {Silvestrini}}]{Gabbiani:1996hi}%
  \BibitemOpen
  \bibfield  {author} {\bibinfo {author} {\bibfnamefont {F.}~\bibnamefont
  {Gabbiani}}, \bibinfo {author} {\bibfnamefont {E.}~\bibnamefont {Gabrielli}},
  \bibinfo {author} {\bibfnamefont {A.}~\bibnamefont {Masiero}}, \ and\
  \bibinfo {author} {\bibfnamefont {L.}~\bibnamefont {Silvestrini}},\
  }\bibfield  {title} {\enquote {\bibinfo {title} {{A Complete analysis of FCNC
  and CP constraints in general SUSY extensions of the standard model}},}\
  }\href {\doibase 10.1016/0550-3213(96)00390-2} {\bibfield  {journal}
  {\bibinfo  {journal} {Nucl. Phys.}\ }\textbf {\bibinfo {volume} {B477}},\
  \bibinfo {pages} {321--352} (\bibinfo {year} {1996})},\ \Eprint
  {http://arxiv.org/abs/hep-ph/9604387} {arXiv:hep-ph/9604387 [hep-ph]}
  \BibitemShut {NoStop}%
\bibitem [{\citenamefont {Kaplan}\ and\ \citenamefont
  {Georgi}(1984)}]{Kaplan:1983fs}%
  \BibitemOpen
  \bibfield  {author} {\bibinfo {author} {\bibfnamefont {David~B.}\
  \bibnamefont {Kaplan}}\ and\ \bibinfo {author} {\bibfnamefont {Howard}\
  \bibnamefont {Georgi}},\ }\bibfield  {title} {\enquote {\bibinfo {title}
  {{SU(2) x U(1) Breaking by Vacuum Misalignment}},}\ }\href {\doibase
  10.1016/0370-2693(84)91177-8} {\bibfield  {journal} {\bibinfo  {journal}
  {Phys.Lett.}\ }\textbf {\bibinfo {volume} {B136}},\ \bibinfo {pages} {183}
  (\bibinfo {year} {1984})}\BibitemShut {NoStop}%
\bibitem [{\citenamefont {Kaplan}\ \emph {et~al.}(1984)\citenamefont {Kaplan},
  \citenamefont {Georgi},\ and\ \citenamefont {Dimopoulos}}]{Kaplan:1983sm}%
  \BibitemOpen
  \bibfield  {author} {\bibinfo {author} {\bibfnamefont {David~B.}\
  \bibnamefont {Kaplan}}, \bibinfo {author} {\bibfnamefont {Howard}\
  \bibnamefont {Georgi}}, \ and\ \bibinfo {author} {\bibfnamefont {Savas}\
  \bibnamefont {Dimopoulos}},\ }\bibfield  {title} {\enquote {\bibinfo {title}
  {{Composite Higgs Scalars}},}\ }\href {\doibase 10.1016/0370-2693(84)91178-X}
  {\bibfield  {journal} {\bibinfo  {journal} {Phys.Lett.}\ }\textbf {\bibinfo
  {volume} {B136}},\ \bibinfo {pages} {187} (\bibinfo {year}
  {1984})}\BibitemShut {NoStop}%
\bibitem [{\citenamefont {Georgi}\ \emph {et~al.}(1984)\citenamefont {Georgi},
  \citenamefont {Kaplan},\ and\ \citenamefont {Galison}}]{Georgi:1984ef}%
  \BibitemOpen
  \bibfield  {author} {\bibinfo {author} {\bibfnamefont {Howard}\ \bibnamefont
  {Georgi}}, \bibinfo {author} {\bibfnamefont {David~B.}\ \bibnamefont
  {Kaplan}}, \ and\ \bibinfo {author} {\bibfnamefont {Peter}\ \bibnamefont
  {Galison}},\ }\bibfield  {title} {\enquote {\bibinfo {title} {{Calculation of
  the Composite Higgs Mass}},}\ }\href {\doibase 10.1016/0370-2693(84)90823-2}
  {\bibfield  {journal} {\bibinfo  {journal} {Phys.Lett.}\ }\textbf {\bibinfo
  {volume} {B143}},\ \bibinfo {pages} {152} (\bibinfo {year}
  {1984})}\BibitemShut {NoStop}%
\bibitem [{\citenamefont {Dugan}\ \emph {et~al.}(1985)\citenamefont {Dugan},
  \citenamefont {Georgi},\ and\ \citenamefont {Kaplan}}]{Dugan:1984hq}%
  \BibitemOpen
  \bibfield  {author} {\bibinfo {author} {\bibfnamefont {Michael~J.}\
  \bibnamefont {Dugan}}, \bibinfo {author} {\bibfnamefont {Howard}\
  \bibnamefont {Georgi}}, \ and\ \bibinfo {author} {\bibfnamefont {David~B.}\
  \bibnamefont {Kaplan}},\ }\bibfield  {title} {\enquote {\bibinfo {title}
  {{Anatomy of a Composite Higgs Model}},}\ }\href {\doibase
  10.1016/0550-3213(85)90221-4} {\bibfield  {journal} {\bibinfo  {journal}
  {Nucl.Phys.}\ }\textbf {\bibinfo {volume} {B254}},\ \bibinfo {pages} {299}
  (\bibinfo {year} {1985})}\BibitemShut {NoStop}%
\bibitem [{\citenamefont {Arkani-Hamed}\ \emph {et~al.}(2002)\citenamefont
  {Arkani-Hamed}, \citenamefont {Cohen}, \citenamefont {Katz},\ and\
  \citenamefont {Nelson}}]{ArkaniHamed:2002qy}%
  \BibitemOpen
  \bibfield  {author} {\bibinfo {author} {\bibfnamefont {N.}~\bibnamefont
  {Arkani-Hamed}}, \bibinfo {author} {\bibfnamefont {A.G.}\ \bibnamefont
  {Cohen}}, \bibinfo {author} {\bibfnamefont {E.}~\bibnamefont {Katz}}, \ and\
  \bibinfo {author} {\bibfnamefont {A.E.}\ \bibnamefont {Nelson}},\ }\bibfield
  {title} {\enquote {\bibinfo {title} {{The Littlest Higgs}},}\ }\href@noop {}
  {\bibfield  {journal} {\bibinfo  {journal} {JHEP}\ }\textbf {\bibinfo
  {volume} {0207}},\ \bibinfo {pages} {034} (\bibinfo {year} {2002})},\ \Eprint
  {http://arxiv.org/abs/hep-ph/0206021} {arXiv:hep-ph/0206021 [hep-ph]}
  \BibitemShut {NoStop}%
\bibitem [{\citenamefont {Aad}\ \emph {et~al.}(2014)\citenamefont {Aad} \emph
  {et~al.}}]{Aad:2014cka}%
  \BibitemOpen
  \bibfield  {author} {\bibinfo {author} {\bibfnamefont {Georges}\ \bibnamefont
  {Aad}} \emph {et~al.} (\bibinfo {collaboration} {ATLAS}),\ }\bibfield
  {title} {\enquote {\bibinfo {title} {{Search for high-mass dilepton
  resonances in pp collisions at $\sqrt{s}=8\,$TeV with the ATLAS detector}},}\
  }\href {\doibase 10.1103/PhysRevD.90.052005} {\bibfield  {journal} {\bibinfo
  {journal} {Phys.Rev.}\ }\textbf {\bibinfo {volume} {D90}},\ \bibinfo {pages}
  {052005} (\bibinfo {year} {2014})},\ \Eprint {http://arxiv.org/abs/1405.4123}
  {arXiv:1405.4123 [hep-ex]} \BibitemShut {NoStop}%
\bibitem [{\citenamefont {Sun}(2014)}]{Sun:2014jha}%
  \BibitemOpen
  \bibfield  {author} {\bibinfo {author} {\bibfnamefont {Sichun}\ \bibnamefont
  {Sun}},\ }\bibfield  {title} {\enquote {\bibinfo {title} {{Little Flavor:
  Heavy Leptons, Z' and Higgs Phenomenology}},}\ }\href@noop {} {\  (\bibinfo
  {year} {2014})},\ \Eprint {http://arxiv.org/abs/1411.0131} {arXiv:1411.0131
  [hep-ph]} \BibitemShut {NoStop}%
\end{thebibliography}%
\end{document}